\documentclass{article}
\usepackage{PRIMEarxiv}
\pdfoutput=1

\usepackage[utf8]{inputenc} 
\usepackage[T1]{fontenc}    
\usepackage{hyperref}       
\usepackage{url}            
\usepackage{booktabs}       
\usepackage{amsfonts}       
\usepackage{nicefrac}       
\usepackage{microtype}      
\usepackage{lipsum}
\usepackage{fancyhdr}       
\usepackage{graphicx}       
\graphicspath{{media/}}     
\usepackage{xcolor}
\usepackage{amsmath,amssymb}

\title{Penguin huddling: a continuum model}

\author{
  Samuel J. Harris and N.R. McDonald \\
  Department of Mathematics \\ 
  University College London \\
  Gower Street , London, WC1E 6BT, UK \\
  sam.harris.16@ucl.ac.uk 
}

\begin{document}
\maketitle

\begin{abstract}
Penguins huddling in a cold wind are represented by a two-dimensional, continuum model. The huddle boundary evolves due to heat loss to the huddle exterior and through the reorganisation of penguins as they seek to regulate their heat production within the huddle. These two heat transfer mechanisms, along with area, or penguin number, conservation, gives a free boundary problem whose dynamics depend on both the dynamics interior and exterior to the huddle. Assuming the huddle shape evolves slowly compared to the advective timescale of the exterior wind, the interior temperature is governed by a Poisson equation and the exterior temperature by the steady advection-diffusion equation. The exterior, advective wind velocity is the gradient of a harmonic, scalar field. The conformal invariance of the exterior governing equations is used to convert the system to a Polubarinova-Galin type equation, with forcing depending on both the interior and exterior temperature gradients at the huddle boundary. The interior Poisson equation is not conformally invariant, so the interior temperature gradient is found numerically using \textcolor{black}{a combined adaptive Antoulas-Anderson and least squares} algorithm. The results show that, irrespective of the starting shape, penguin huddles evolve into an egg-like steady shape. This shape is dependent  on the wind strength, parameterised by the P\'eclet number Pe, and a parameter $\beta$ which effectively measures the strength of the interior self-generation of heat by the penguins. The numerical method developed is applicable to a further five free boundary problems.
\end{abstract}


\section{Introduction}\label{sec:intro}

Among all the penguin species living in the Antarctic region \cite{le1977emperor}, the emperor penguin (Aptenodytes forsteri) experiences the most extreme weather conditions, enduring wind speeds over $30$ms$^{-1}$ at temperatures below $-40^\circ$C \cite{mccafferty2013emperor,le1977emperor,williams2015hidden,gerum2013origin}. Unlike other species such as the gentoo and the Ad\'elie \cite{kooyman1976heat}, emperor penguins have no fixed nest \cite{waters2012modeling,gerum2018structural}, allowing the birds to huddle together for warmth in severe wind conditions. This strategy is very effective; ambient temperatures within the huddle often exceed 20$^\circ$C, with a maximum temperature of 37.5$^\circ$C having been recorded \cite{gilbert2006huddling}. Such huddling is crucial for the penguins' survival, especially during foraging trips \cite{kirkwood1999occurrence} and throughout their breeding season, which coincides with the Antarctic winter \cite{williams2015hidden}.

Over the few hours that huddling events typically last \cite{gilbert2006huddling,waters2012modeling}, the penguins are continuously (albeit slowly) on the move. The birds constantly reorganise their position in the huddle, where \cite{le1977emperor} observes that ``birds that at first are in the center of the huddle become members of the rear flank [the side most exposed to the wind] and move, in their turn, up the sideline". This process benefits all members of the huddle and allows the heat to be equally shared: those at the edge do not remain cold and those at the centre do not overheat. These individual penguin movements affect the propagation of the huddle as a whole - the huddle is seen to propagate downwind \cite{ancel2015new} and the shape traced out by the edge of the huddle, or huddle boundary, may change over time \cite{gerum2013origin,waters2012modeling}. It is the propagation and evolution of the huddle boundary shape that is the focus of this work.

The modelling of penguin huddle dynamics has received increased attention over the last decade, but still remains sparse in the literature, despite numerous observations and findings from the field, see \cite{ancel2015new,mccafferty2013emperor,williams2015hidden,le1977emperor}. The works \cite{zitterbart2011coordinated, gerum2013origin,gerum2018structural} draw the analogy between penguin huddle/colony formation and condensed matter physics, with Lennard-Jones-like forces used to represent attractive and repulsive interactions between penguins that allow complex, lattice structures (such as  huddles) to form. This idea is expanded upon in \cite{mina2018penguin}, where Gaussian processes are used to test if robots could adopt this penguin inspired huddling strategy. Similarly, \cite{richter2018phase} liken the rearrangement of the penguin huddle to a phase transition from solid (dense huddle) to liquid (loose huddle) to gas (no huddle) depending on environmental factors such as temperature and wind speed. A rigorous fluid dynamics approach is taken in \cite{gu2018novel}, where a finite difference method is used to solve the full Navier-Stokes equations of the surrounding fluid (the exterior wind), giving a two-way coupling between the environment and the behaviour of the huddle. Using a simpler fluids based model, \cite{waters2012modeling} assume the exterior wind to be a two-dimensional, irrotational flow of an incompressible fluid and from this model, the temperature profile exterior to the huddle and individual penguin movements can be found.

In this work, the approach of \cite{waters2012modeling} is expanded upon with two key differences. First, a simple model for the effect of interior penguin reorganisation as a result of self-generation of heat throughout the entire huddle, as observed in \cite{le1977emperor}, is included. In \cite{waters2012modeling}, while penguin reorganisation around the huddle boundary was acknowledged, it was assumed that interior penguins were so tightly packed that they could not change their position in the huddle. That assumption is here dropped, yet it is still assumed that the penguins are packed tightly enough such that wind does not permeate through the huddle and so the wind flow remains entirely in the exterior region. Throughout this work, the phrases ``exterior only penguin problem" and ``full penguin problem" are used to refer to the penguin huddle problem without and with interior penguin reorganisation, respectively. Second, the penguin huddle itself is treated as a continuum. Existing huddle models treat penguins as discrete elements (or particles) such that individual penguin movements can easily be tracked. For sufficiently large huddles on the order of a thousand penguins \cite{le1977emperor}, these discrete models quickly become computationally expensive. Continuum models, as used here, are often faster, cheaper, and preferred when studying the change in shape and position of the overall huddle.

Using a continuum to model the interior and exterior of the full penguin problem draws analogues to other two-dimensional free boundary problems. Foremost among these is the classical Hele-Shaw free-boundary problem and its extensions, including  Saffman-Taylor fingering (a consequence of injecting a fluid into another, more viscous fluid) \cite{saffman1958penetration,howison1986fingering,paterson1981radial}, the expansion and contraction of bubbles in a Hele-Shaw cell \cite{taylor1959note,entov1991bubble,dallaston2013accurate,dallaston2016curve} and the dissolution of soluble objects in two-dimensional potential flow \cite{ladd2020}. Using conformal mapping to reformulate the free boundary problem as a Polubarinova-Galin (PG) equation has proved effective in finding exact and numerical solutions for this class of problems. Solidification and melting problems have also been extensively studied using analogous techniques, see \cite{cummings1999two,rycroft2016asymmetric,mullins1964stability,choi2005steady,tsai2007star,grodzki2019reactive} and \cite{goldstein1978effect}; the latter of these examines the freezing of porous media flow by freeze pipes, with the velocity of the free boundary dependent on both interior and exterior temperature gradients, similar to the full penguin problem. Free boundary problems arise in other natural processes, including, but not limited to, dendritic crystal growth \cite{langer1980instabilities,brower1984geometrical} and flame front propagation \cite{sethian1985curvature,hilton2016curvature,harris2022fingering}. Continua may also be used to \textcolor{black}{model other collective behaviours \cite{gerum2013origin,burger2013individual,bernardi2020agent}} such as fish schooling \cite{katz2011inferring}, bird flocking \cite{bhattacharya2010collective}, bat swarming \cite{herreid1963temperature,ryan2019changes} and ant colony formation \cite{nave2020attraction,ko2022fire}.

There are two  heat transfer mechanisms in the full penguin problem, which formulate a suitable boundary condition for the propagation of the huddle boundary. The first is the cooling effect of the wind on the huddle. As in \cite{waters2012modeling}, the incompressible wind flow is assumed irrotational and flowing sufficiently faster than the propagation of the huddle, thus, the flow can be treated as steady. As such, the wind flow and temperature transport in the region exterior to the penguin huddle can be modelled by the Laplace and the steady advection-diffusion equations, respectively. This coupled pair of partial differential equations (PDEs) is conformally invariant \cite{bazant2005conformal} - a property which motivates the use of a conformal mapping approach in the numerical method \cite{dallaston2016curve,cummings1999two,ladd2020,rycroft2016asymmetric,harris2022fingering}. The exterior region and the huddle boundary in the physical $z-$plane can be conformally mapped to the exterior and boundary of the unit disk in some ``mathematical" (or ``canonical") $\zeta-$plane, and vice versa. Therefore, the physical coordinates can be expressed in terms of the conformal map $z=f(\zeta,t)$ and the condition on the huddle boundary rewritten as a PG type equation \cite{gustafsson2006conformal}. The task then becomes that of finding the conformal map.

The second heat transfer mechanism is that of the uniform, self-generation of heat by the penguins in the interior. Its steady diffusion amongst the densely packed penguins is modelled by a Poisson equation for the interior temperature with a constant forcing term. By the Riemann mapping theorem (see e.g. \cite{bazant2005conformal}), the interior of the penguin huddle will not also be mapped to the interior of the $\zeta-$disk by the conformal map $z=f(\zeta,t)$ used to map the exteriors. Further, the Poisson equation is not conformally invariant \cite{mcdonald2015poisson}. Hence, to solve for temperature in the interior using a conformal mapping approach, a different conformal map $f_2$ would need to be found numerically and the transformed Poisson equation in the $\zeta-$ plane solved. Alternatively, the interior temperature can be found directly in the $z-$plane without the use of conformal mapping. Here, the methods developed by Trefethen and colleagues \cite{trefethen2018series,trefethen2020numerical,baddoo2020lightning,costa2020solving,costa2021aaa} are used to find a rational approximation of a harmonic function in some simply connected domain as the real part of some analytic function, subject to given boundary conditions. This function consists of a power series in $z$ and a series of rational functions involving poles in the exterior clustered near singularity points \cite{trefethen2020numerical}. Both the adaptive Antoulas-Anderson algorithm (AAA, pronounced triple-A) \cite{costa2020solving} and a least-squares (LS) algorithm, such as the lightning Laplace solver \cite{baddoo2020lightning} can be used to numerically find this analytic function \cite{costa2021aaa} and hence the required harmonic function. In this work, by first finding a particular solution, the problem of solving the interior Poisson equation then becomes one of solving the Laplace equation with a modified boundary condition. In turn, this is readily solved numerically by the AAA-LS method.

The structure of this paper is as follows. The two-dimensional, continuum penguin huddle model is formulated in Sect. \ref{sec:model}, with assumptions stated and dimensionless quantities introduced. A suitable boundary condition for the normal boundary velocity $v_n$ is derived, incorporating effects of exterior wind cooling, internal heat generation by the penguins and area (penguin) conservation.  The numerical method is then developed in Sect. \ref{sec:num}, where the equation for $v_n$ is rewritten as a PG type equation using a conformal mapping approach. Attention here is given to  the exterior heat flux $\sigma$ in the $\zeta-$plane and the interior heat flux which is found in the physical plane. Results are presented in Sect. \ref{sec:results}, illustrating the evolution of penguin huddles of different starting shapes under a variety of wind and interior heating effect strengths. Conclusions  are  discussed in Sect. \ref{sec:discuss} and extensions to the work, and applications to further (free boundary) problems, noted. 

\section{Model setup}\label{sec:model}

\begin{figure}[h]%
\centering
\includegraphics[width=0.95\textwidth]{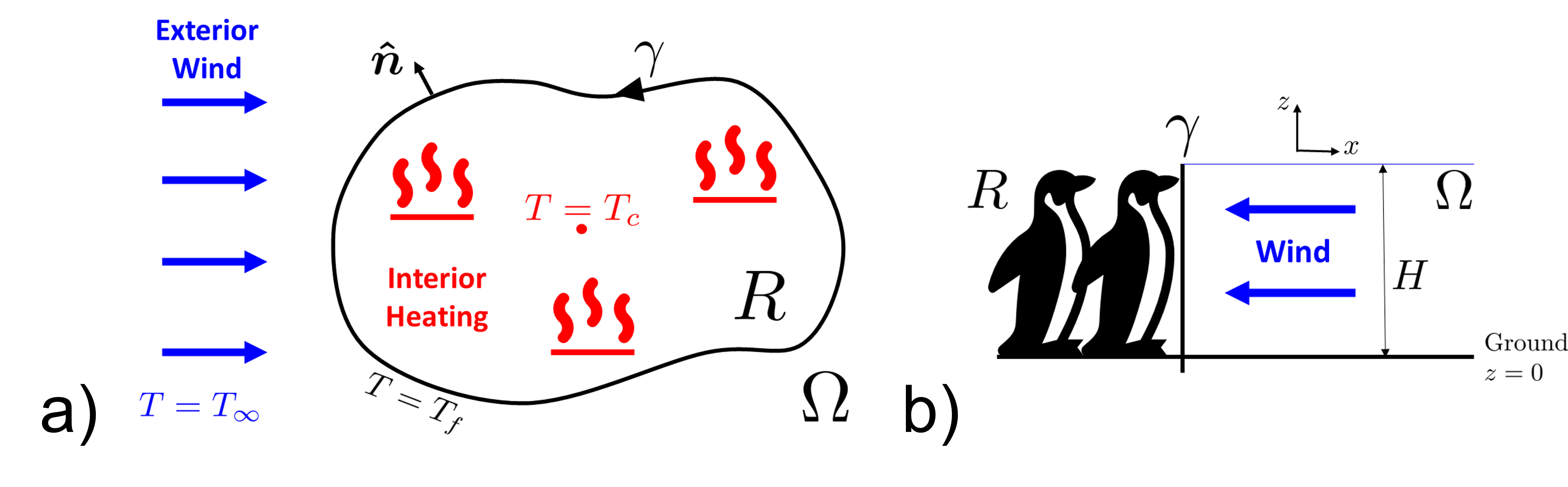}
\caption{Schematic diagram of the full penguin problem showing the interior region $R$ and exterior region $\Omega$. a) Plan view where $T_c$ denotes the maximum temperature $T$ inside the huddle, b) side view.}\label{fig:model}
\end{figure}

Consider a huddle of penguins with horizontal length scale $L$ where the penguins are assumed homogeneous in size and shape. \textcolor{black}{The huddle is considered as a continuum with penguins packed as tightly as possible while still allowing for huddle reorganisation \cite{zitterbart2011coordinated} and hence uniform spacing between each penguin can be assumed constant across the entire huddle and for all time.} Emperor penguins have a typical height of $H=1m$ \cite{le1977emperor} and their tight packing means that background winds are diverted around the huddle, allowing  the dynamics to be considered as two-dimensional, see Figure \ref{fig:model}. The boundary of the penguin huddle $\gamma(t)$ is defined as the two-dimensional, non-overlapping curve enclosing the huddle; the finite\textcolor{black}{, simply connected} interior of the curve occupied by the huddling penguins is labelled $R$ and the penguin-free exterior as $\Omega$. The boundary curve $\gamma$ is orientated anti-clockwise such that the unit normal $\boldsymbol{\hat{n}}$ points from the interior to the exterior.

The huddle $R$ is buffeted by \textcolor{black}{a unidirectional and constant strength wind} of magnitude $U$ which, without loss of generality, flows in the positive $\boldsymbol{\hat{x}}$ direction.  The wind is the irrotational flow of air, which is considered to be an \textcolor{black}{inviscid} and incompressible fluid which flows sufficiently rapidly compared to the evolution of the penguin huddle as a whole, so that the transport of temperature can be treated as steady \cite{waters2012modeling}. \textcolor{black}{In reality, viscous effects taking the form of thin boundary layers around the exterior of the penguin huddle are present. In this work, these are ignored in order to facilitate the accurate computation of the exterior potential flow using conformal mapping based methods.} Denote the temperature far from the huddle as $T_\infty$, the temperature on the huddle boundary as $T_f$ and the maximum temperature in the huddle interior as $T_c$.

The equations governing the wind and exterior temperature transport are, respectively, the Laplace equation and the steady advection-diffusion equation
\begin{gather}
    \nabla^2\phi=0\;\;\;\;\text{in }\Omega,\label{eq:potflow1}\\
    \boldsymbol{u\cdot\nabla}T_\Omega = D_\Omega\nabla^2T_\Omega \;\;\;\;\text{in }\Omega,\label{eq:staddif1}
\end{gather}
where $\phi$ is the wind velocity potential, $\boldsymbol{u}=\boldsymbol{\nabla}\phi$ is the wind velocity, $T_\Omega$ is the temperature in $\Omega$ and $D_\Omega$ is the diffusivity of temperature in air.  The associated boundary and far field conditions are 
\begin{gather}
    T_\Omega=T_f,\;\;\frac{\partial\phi}{\partial n}=0\;\;\;\;\text{on }\gamma,\label{eq:extbc1}\\
    T_\Omega\rightarrow T_\infty,\;\;\phi\rightarrow Ux\;\;\;\;\text{as }r\rightarrow\infty\label{eq:extff1}.
\end{gather}

Now, consider the interior $R$ of the huddle \textcolor{black}{where there is no wind. Owing to their ability to self-generate heat, which subsequently diffuses within the huddle, penguins nearer the huddle centre are warmer than those at the boundary \cite{le1977emperor,zitterbart2011coordinated}, i.e. $T_c>T_f$. This source of heat provides a further contribution to the heat flux at the boundary $\gamma$. Consistent with the continuum assumption, it is assumed that heat is generated uniformly across the huddle. Further, as in the exterior, it is assumed the interior heat diffuses quickly compared to the evolution of the boundary. Thus the temperature distribution in $R$ is modelled by the steady heat equation }
\begin{equation}\label{eq:intheat1}
    \nabla^2T_R=-\frac{Q}{D_R}\;\;\;\;\text{in }R,
\end{equation}
where $D_R$ is the diffusivity of the interior temperature $T_R$ through the penguin huddle and $Q$ is the (constant) \textcolor{black}{source} term representing the \textcolor{black}{self-generation of heat by} the penguins. 

In order to obtain parameter estimates for the interior problem, it is helpful to first consider a  circular penguin huddle of radius $L$, noting that $T_R(0)=T_c$. \textcolor{black}{ The Poisson} equation \eqref{eq:intheat1} can then be solved exactly, giving
\begin{subequations}\label{eq:Q/D}
\begin{align}
    T_R = T(r) &= T_c-\frac{Q}{4D_R}r^2, & \text{where }\frac{Q}{D_R}&=\frac{4(T_c-T_f)}{L^2}.
\tag{\theequation a,b}
\end{align}
\end{subequations}
The relation (\ref{eq:Q/D}\textcolor{black}{b}) is used in Sect. \ref{subsec:nondim} to non-dimensionalise the problem for cases when the region $R$ is non-circular.

\textcolor{black}{Finally, a condition on $\gamma(t)$ is sought by specifying its outward normal velocity $v_n$. Following \cite{waters2012modeling}, the idea that the boundary evolves owing to the heat loss of penguins on the boundary is used. Consequently, the problem has some similarity to the two-phase Stefan problem \cite{cummings1999two,goldstein1978effect,rycroft2016asymmetric,gupta2017classical} with the boundary temperature gradient driving its evolution, yet, unlike such problems, no phase change occurs. Waters {\it et al.} \cite{waters2012modeling} move individual boundary penguins according to their heat loss to the exterior, whereas here both exterior and interior  heat fluxes are considered. First, the heat loss to the exterior wind is the flux $\kappa_\Omega\boldsymbol{\hat{n}\cdot\nabla}T_\Omega$ evaluated on $\gamma(t)$, where $\kappa_\Omega$ is the thermal conductivity of air.  Second, the heat gain due to the warm huddle interior is $-\kappa_R\boldsymbol{\hat{n}\cdot\nabla}T_R$ evaluated on $\gamma(t)$, where $\kappa_R$ is the thermal conductivity of huddling penguins. Equating the normal velocity with the net heat flux gives
\begin{equation}\label{eq:energycons}
    Av_n = \kappa_\Omega\boldsymbol{\hat{n}\cdot\nabla}T_\Omega-\kappa_R\boldsymbol{\hat{n}\cdot\nabla}T_R,\quad{\rm on}\quad \gamma(t),
\end{equation}
where the constant term $A$ is chosen to ensure a match in dimensions between both sides of \eqref{eq:energycons}, with the particular choice of $A$ affecting the timescale for the boundary evolution.  For convenience in the non-dimensionalisation process upcoming in Sect. \ref{subsec:nondim}, let $A = \lambda\rho_\Omega c_\Omega (T_f-T_\infty)$, where $\rho_\Omega$ and $c_\Omega$ are the density and specific heat capacity of the air, respectively, and $\lambda$ is a non-dimensional constant to be chosen later.}

\textcolor{black}{
The right hand side of \eqref{eq:energycons} is the jump in heat flux at the boundary. This observation reinforces the analogy with the Stefan problem where typically the normal velocity of the boundary depends on the jump in the heat flux  \cite{gupta2017classical}. For example, analogous boundary conditions arise in other natural free boundary problems, including the dissolution of finite objects, where only the gradient of the agent undergoing advection and diffusion exterior to the object is important  \cite{rycroft2016asymmetric,ladd2020}, or in porous media flow about freeze pipes, in which the temperature gradient in both the exterior fluid and interior ice regions are significant \cite{goldstein1978effect}. However it is emphasised that, in this work, \eqref{eq:energycons} is purely a modelling assumption.}

\textcolor{black}{Rewriting \eqref{eq:energycons} gives the following condition on $\gamma$
\begin{equation}\label{eq:vn1}
    \lambda\rho_\Omega c_\Omega (T_f-T_\infty) \boldsymbol{\hat{n}\cdot}\frac{\partial\boldsymbol{x}}{\partial t} = \kappa_\Omega\boldsymbol{\hat{n}\cdot\nabla}T_\Omega-\kappa_R\boldsymbol{\hat{n}\cdot\nabla}T_R + \overline{C}(t).
\end{equation}
 The additional term $\overline{C}$(t) in \eqref{eq:vn1} is needed to enforce conservation of penguins, or, equivalently, area conservation \cite{waters2012modeling}, namely that
\begin{equation}\label{eq:areacons}
    \int_\gamma \boldsymbol{\hat{n}\cdot}\frac{\partial\boldsymbol{x}}{\partial t} \text{d}s=\int_\gamma v_n \text{d}s = 0.
\end{equation}
}

\subsection{Non-dimensionalisation and parameter values}\label{subsec:nondim}
In order to identify the  parameters of the problem, the model \eqref{eq:potflow1}-\eqref{eq:intheat1}, \eqref{eq:vn1} is non-dimensionalised, with  non-dimensional quantities temporarily labelled as starred variables. Let $L$ characterise the length scale of region $R$  i.e. $\boldsymbol{x} = L\boldsymbol{x}^\ast$, and use the far-field wind magnitude $U$ as the characteristic velocity scale, $\boldsymbol{u} = U\boldsymbol{u}^\ast$. The following scalings are also used
\begin{subequations}
\begin{equation}
    \boldsymbol{\nabla} =\frac{1}{L}\boldsymbol{\nabla}^\ast, \;\;\; T^\ast = \frac{T-T_\infty}{T_f-T_\infty},\;\;\; t = \frac{1}{\tau}t^\ast,\;\;\;\tau = \frac{U}{L}.
\tag{\theequation a,b,c,d}
\end{equation}
\end{subequations}

 Using (\ref{eq:Q/D}\textcolor{black}{b}), the non-dimensional form of the Poisson equation \eqref{eq:intheat1} can be written (now dropping the stars) as
\begin{subequations}\label{eq:intheat2}
\begin{equation}
    \nabla^2T_R=-\alpha,\;\;\;\;\;\; \alpha=\frac{4(T_c-T_f)}{T_f-T_\infty}.
\tag{\theequation a,b}
\end{equation}
\end{subequations}
The boundary condition \eqref{eq:vn1} becomes
\begin{equation}\label{eq:vn2}
    \lambda\text{Pe } \boldsymbol{\hat{n}\cdot}\frac{\partial\boldsymbol{x}}{\partial t} = \boldsymbol{\hat{n}\cdot\nabla}T_\Omega-\beta\boldsymbol{\hat{n}\cdot\nabla}T_R + C(t),
\end{equation}
where $\text{Pe}=UL/D_\Omega$ is the P\'eclet number, with $D_\Omega = \kappa_\Omega/\rho_\Omega c_\Omega$ the thermal diffusivity of air, $\beta=\kappa_R/\kappa_\Omega$ is the ratio of thermal conductivities and $C(t)$ is the dimensionless area conserving term, an explicit expression for which can be found by substituting \eqref{eq:vn2} into \eqref{eq:areacons} (see \eqref{Meq:C} below). \textcolor{black}{Note that the appearance of Pe on the left hand side of \eqref{eq:vn2} affects the time scaling reported in the results. The real world dimensional time scale ``$t_r$" is therefore $t_r=\lambda\text{Pe}\;(L/U)t^{\ast}$. In what follows,  the choice $\lambda=1$ is made and justified {\it a posteriori} by comparing the time for the huddle to evolve with observations - see Sect. \ref{sec:discuss}.}

There are three parameters: $\text{Pe}, \beta$ and $\alpha$. Following the justification in \cite{waters2012modeling}, the wind  around the penguin huddle is likely turbulent and hence the P\'eclet number can be approximated by a turbulent Reynolds number, which is $\mathcal{O}(100)$ for Antarctic winds at this scale. Consistent with \cite{waters2012modeling}, \textcolor{black}{the present work considers P\'eclet values between $1$ and $100$.} The values of the remaining two parameters, $\alpha$ and $\beta$, are estimated as follows. The air temperature far upstream is taken to be $T_\infty=-40^\circ$C \cite{mccafferty2013emperor,williams2015hidden} while ambient temperatures in the huddle are above freezing \cite{zitterbart2011coordinated}. It seems reasonable then to let $T_f=0^\circ$C. The centre of a huddle can exceed $20^\circ$C, with a maximum temperature of $37.5^\circ$C having been recorded \cite{gilbert2006huddling}. Using, therefore,  $T_c=30^\circ$C gives $\alpha=3$. 

From table C.5 of \cite{basu2018biomass}, the thermal conductivity of air at $-40^\circ$C is approximately $\kappa_\Omega\approx0.022Wm^{-1}K^{-1}$. The thermal conductance of penguins is $1.3Wm^{-2}K^{-1}$ from \cite{le1977emperor}, however this is not the same as the thermal conductivity - it has different units for example. To find the thermal conductivity, the conductance must be multiplied by the thickness of the penguin insulation layer which is $d=0.024m$ \cite{dawson1999heat}. Therefore, $\kappa_R=0.0312Wm^{-1}K^{-1}$ \textcolor{black}{and so $\beta=1.4182\approx1.5=\mathcal{O}(1)$. This estimate for $\beta$ depends on the wind flow being either laminar or turbulent, with $\beta$ decreasing as the flow becomes more turbulent and the effective thermal conductivity of air increases \cite{kittel1996introduction}. In this work, a range of $\beta$ values between $\beta=0$ and $\beta=1.5$  are considered with $\beta=0.14\approx0.1=\mathcal{O}(10^{-1})$ taken to be a representative value.}

To summarise, the full dimensionless system for the penguin problem is
\begin{subequations}
\begin{gather}
    \text{Pe } v_n = \boldsymbol{\hat{n}\cdot\nabla}T_\Omega-\beta\boldsymbol{\hat{n}\cdot\nabla}T_R+C(t),\;\;\;\text{on }\gamma,\label{Meq:vn}\\
    \nabla^2\phi=0\;\;\;\;\text{in }\Omega,\label{Meq:potflow}\\
    \text{Pe }\boldsymbol{u\cdot\nabla}T_\Omega = \nabla^2T_\Omega \;\;\;\;\text{in }\Omega,\label{Meq:addiff}\\
    \nabla^2T_R=-\alpha\;\;\;\;\text{in }R,\label{Meq:Poisson}\\
    T_\Omega=T_R=1,\;\;\frac{\partial\phi}{\partial n}=0\;\;\;\;\text{on }\gamma,\label{Meq:bdryc}\\
    T_\Omega\rightarrow 0,\;\;\phi\rightarrow x\;\;\;\;\text{as }r\rightarrow\infty,\label{Meq:farfc}\\
    C(t) = \frac{-\int_\gamma (\boldsymbol{\hat{n}\cdot\nabla}T_\Omega-\beta\boldsymbol{\hat{n}\cdot\nabla}T_R) \text{d}s}{\int_\gamma \text{d}s}. \label{Meq:C}
\end{gather}
\end{subequations}
 The task is to solve the system \eqref{Meq:vn}--\eqref{Meq:C} and, in particular, find the evolution of the $\gamma(t)$ for some given initial huddle shape $\gamma(0)$. \textcolor{black}{This is done in Sect. \ref{sec:results} with the following choices: $\text{Pe}=1\to 100$, $\beta=0\to 1.5$} and $\alpha=3$.

 \section{Numerical method}\label{sec:num}
Following the work of \cite{goldstein1978effect,ladd2020}, the free boundary problem \eqref{Meq:vn}--\eqref{Meq:C} can be written as a PG type equation \cite{bazant2005conformal}  in terms of a conformal map $z=f(\zeta,t)$ from the exterior of the unit disk in the canonical $\zeta-$plane to the exterior region $\Omega$ in the physical $z-$plane \cite{choi2005steady,waters2012modeling}. However, the interior $R$ is not mapped using $f(\zeta,t)$, thus care must be taken with the term involving the interior temperature $T_R$.

First, note the following which hold on the unit disk $r=\lvert\zeta\vert=1$   \cite{dallaston2013accurate,ladd2020}
\begin{subequations}\label{eq:PQexpr}
\begin{equation}
v_n=\frac{\text{Re}\Big[f_t\overline{\zeta f_\zeta}\Big]}{\lvert f_\zeta\vert},\;\;\;\;
\boldsymbol{\hat{n}\cdot\nabla}T_\Omega = \text{Re}[n\overline{\nabla}T_\Omega]=\frac{1}{\lvert f_\zeta\vert}\frac{\partial T_\Omega}{\partial r} = \frac{\sigma(\theta)}{\lvert f_\zeta\vert},
\tag{\theequation a,b}
\end{equation}
\end{subequations}
where $\overline{\nabla}=\partial_x -i\partial_y$, $\zeta=e^{i\theta}$, $f_\zeta = \partial f/\partial\zeta$ and $n=n_x+in_y=\zeta f_\zeta/\lvert f_\zeta\vert$ is the complex representation of the normal vector in the $z-$plane \cite{harris2022fingering}. The function $\sigma(\theta)$ is found by solving the following integral equation \cite{goldstein1978effect,choi2005steady,ladd2020}
\begin{equation}\label{eq:inteq}
    \pi = \int_0^\pi K\big[\text{Pe}(\cos{\theta}-\cos{\theta'})\big]\sigma(\theta')\text{d}\theta',
\end{equation}
where $K(x)=e^x K_0(\lvert x\vert)$ and $K_0$ is the modified Bessel function. The function $\sigma$ is the heat flux (due to the exterior effect alone) around the unit $\zeta-$disk \cite{ladd2020}. A numerical solution to \eqref{eq:inteq} can be found at $M$ equally spaced points on the upper half of the unit disk $\theta\in(0,\pi)$, see appendix A of \cite{ladd2020}. To evaluate $\sigma$ on the remainder of the disk, note that $\sigma(\theta^\ast)=\sigma(\pi+\theta^\ast)$, where $0\leq\theta^\ast<\pi$, and $\sigma(\theta)\rightarrow 1/\pi$ as $\theta\rightarrow0$ \cite{choi2005steady,ladd2020}.

Second, consider the contribution from the interior $\boldsymbol{\hat{n}\cdot\nabla}T_R$. This is found by direct computation in the $z-$plane. To begin, introduce the function $u=T_R+\alpha r^2/4$; from \eqref{Meq:Poisson} and \eqref{Meq:bdryc}, $u$ satisfies
\begin{gather}
    \nabla^2u=0 \;\;\;\;\text{in }R,\\
    u=1+\frac{\alpha r^2}{4} \;\;\;\;\text{on }\gamma.
\end{gather}
This is a Laplace problem in the interior of a smooth, simply connected domain with  Dirichlet boundary condition $u=h(z)$ on $\gamma$. Efficient methods developed by Trefethen and colleagues are here used to solve the Laplace problem directly, see \cite{gopal2019solving,trefethen2020numerical} for further details. These are based on a rational approximation of $u$ in the complex $z-$plane: there exists an analytic function $F(z)$ such that $u=\text{Re}[F(z)]$ which can be approximated by the series solution
\begin{equation}\label{eq:F(z)}
    F(z) = A_0+\sum_{k=1}^{N_1}A_k(z-c_0)^k+\sum_{k=1}^{N_2}\frac{B_k}{z-p_k},
\end{equation}
where $A_0, A_k$ and $B_k$ are constants, $N_1$ is the power series truncation, $c_0$ is a point in the interior $R$ (in this work, the conformal centre of the polygon is used \cite{ladd2020}) and $p_k$ are some $N_2$ poles in the exterior $\Omega$. The first sum is known as the ``Runge" (or smooth) part and the second as the ``Newman" (or singular) part \cite{gopal2019solving,costa2020solving}. 

Suitable poles $p_k$ are found by the ``adaptive Antoulas-Anderson" (AAA, pronounced `triple-A') algorithm \cite{nakatsukasa2018aaa,costa2020solving} and the unknown coefficients $A_0, A_k$ and $B_k$ by a least-squares (LS) algorithm \cite{trefethen2018series,baddoo2020lightning,brubeck2021vandermonde}: combining these into a AAA-LS method \cite{costa2021aaa} can thus find $F(z)$ numerically. It follows \cite{trefethen2018series} that
\begin{equation}
    \nabla u = \nabla\text{Re}[F(z)]=\overline{F'(z)},
\end{equation}
where the primed notation indicates differentiation, in this case with respect to $z$. Therefore,
\begin{equation}
    \overline{\nabla}T_R = \overline{\nabla}[u-\alpha r^2/4]=F'(z) - \frac{\alpha}{2}\overline{z},
\end{equation}
recalling that $\overline{\nabla}=2\partial_z$ and $r^2=z\overline{z}$. This gives
\begin{equation}\label{eq:intexpr}
    \boldsymbol{\hat{n}\cdot\nabla}T_R = \text{Re}[n\overline{\nabla}T_R]=\frac{1}{\lvert f_\zeta \vert}\text{Re}[\zeta f_\zeta(F'(z)-\alpha\overline{z}/2)]\equiv\frac{\omega(\theta)}{\lvert f_\zeta \vert},
\end{equation}
where the boundary points $z$ in \eqref{eq:intexpr} are considered in the $\zeta-$plane so that the right hand side of \eqref{eq:intexpr} is purely a function of $\theta$. Drawing an analogue with $\sigma$, the function $-\beta\omega$ is the heat flux around the unit $\zeta-$disk due to the interior effect alone.

Third, the formula \eqref{Meq:C} for the area conserving term $C(t)$ can be transformed to the $\zeta-$plane, noting that d$s=\lvert f_\zeta\vert\text{d}\theta$. 
Substituting (\ref{eq:PQexpr}\textcolor{black}{b}) and \eqref{eq:intexpr} into \eqref{Meq:C} gives
\begin{equation}\label{eq:Cexpr}
    C(t)= \frac{-\int_{-\pi}^{\pi}\big(\sigma-\beta\omega\big)\text{d}\theta}{\int_{-\pi}^{\pi} \lvert f_\zeta\vert\text{d}\theta}.
\end{equation}
Therefore, using (\ref{eq:PQexpr}\textcolor{black}{a,b}), \eqref{eq:intexpr} and \eqref{eq:Cexpr}, the PG form of \eqref{Meq:vn} is
\begin{equation}\label{Meq:PGvn}
    \text{Pe Re}[f_t\overline{\zeta f_\zeta}] = \sigma(\theta)-\beta\omega(\theta)+C(t)\lvert f_\zeta\vert,
\end{equation}
where $\sigma(\theta)$ satisfies \eqref{eq:inteq} and $\omega(\theta)=\text{Re}[\zeta f_\zeta(F'(z)-\alpha\overline{z}/2)]$, with $F(z)$ given by \eqref{eq:F(z)}. Note that $\sigma(\theta)-\beta\omega(\theta)$ is the total heat flux across  $\lvert\zeta\rvert=1$; recall that the heat flux across the huddle boundary $\gamma$ is $\boldsymbol{\hat{n}\cdot\nabla}T_\Omega-\beta\boldsymbol{\hat{n}\cdot\nabla}T_R$. 

It remains to find the conformal map $f$, the general form of which is \cite{rycroft2016asymmetric}  
\begin{equation}\label{eq:cmap}
    z=f(\zeta,t)= a_{-1}(t)\zeta + \sum_{k=0}^\infty c_k(t)\zeta^{-k},
\end{equation}
where $a_{-1}(t)$ is a real function in time and $c_k(t)=a_k(t)+ib_k(t)$ are complex functions in time. Note that $a_{-1}$ is the conformal radius and $c_0$ is the conformal centre \cite{ladd2020}. The same $c_0$  is used as the interior point in constructing the power series solution \eqref{eq:F(z)} for the Poisson problem in $R$. Numerically, the Laurent series \eqref{eq:cmap} is truncated at $N$ terms, giving $n=2(N+1)+1=2N+3$ unknown real functions in time: $a_{-1}(t)$, $a_k(t),b_k(t)$, $k=0,1,\cdots,N$. Likewise, $n$ equally spaced points around the unit $\zeta-$disk are chosen; the first point is at $\theta=0$, the next $N+1$ points are in the interval $\theta\in(0,\pi)$ and the final $N+1$ points in $\theta\in(\pi,2\pi)$. Hence the choice $M=N+1$ in $(0,\pi)$ is made when numerically solving the integral equation \eqref{eq:inteq} for $\sigma(\theta)$. The PG equation \eqref{Meq:PGvn} becomes a system of $n$ coupled ordinary differential equations (ODEs)  determining the time evolution of the $n$ Laurent coefficients (see \cite{dallaston2013accurate,dallaston2016curve,harris2022fingering}). In this work, the MATLAB routine $ode15i$ was used to solve this ODE system. 

\section{Results}\label{sec:results}
In the results presented here, the evolution of the huddle boundary is plotted at equal time intervals in the range $[0,t_{max}]$, where $t_{max}$ is the maximum time. The series truncation $N=128$ is chosen, leading to a system of $n=259$ ODEs to be solved. The initial starting shape of the penguin huddle is taken to be either a circle, a slanted ellipse, a triangle, an irregular pentagon \textcolor{black}{or an hourglass}. The initial Laurent coefficients $a_{-1}(0)$, $c_k(0)$ for the conformal map \eqref{eq:cmap} of these shapes are taken from \cite{rycroft2016asymmetric} (see their Figures 2a, 2b, 2d, 5a \textcolor{black}{and 7a}, respectively) and scaled such that each shape encloses the same area\footnote{This scaling can be found analytically using the complex form of Green's theorem \cite{mcdonald2015poisson} to find the exact area from the conformal map $z=f(\zeta,0)$.}. Area conservation of the huddle was monitored during each numerical experiment and found to hold with a relative error of $\mathcal{O}(10^{-5})$. As a test, experiments were run which neglected the interior (i.e. $\beta=0$) and the area conserving ($C(t)$) terms; the geometry of the obtained solutions matched quantitatively with results in \cite{ladd2020} who considered an analogous problem in this class.

\begin{figure}[h]%
\centering
\includegraphics[width=0.95\textwidth]{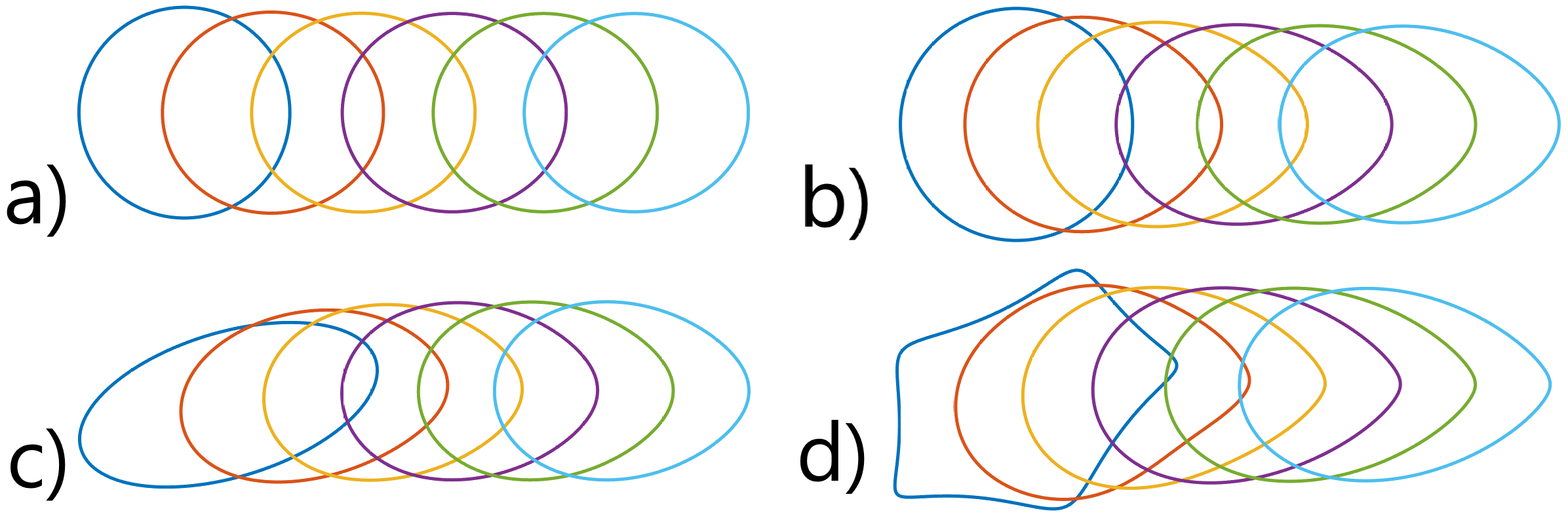}
\caption{Exterior only ($\beta=0$) penguin problem. Shapes are plotted at equal time intervals up to a maximum time $t_{max}$. \textcolor{black}{The $x$ and $y$ axes (not shown) are scaled at a 1-to-1 aspect ratio.} Each description refers to the initial penguin huddle shape, the P\'eclet number and the value of $t_{max}$: a) circle with Pe $=1$, $t_{max}=10$, b) circle with Pe $=10$, $t_{max}=20$, c) slanted ellipse with Pe $=10$, $t_{max}=20$, d) irregular pentagon with Pe $=100$, $t_{max}=50$.}\label{fig:waters}
\end{figure}

First, consider the ``exterior only" penguin problem when $\beta=0$. This is a continuum analog of the discrete model developed in \cite{waters2012modeling}, where the penguin huddle evolves in response to the wind but the reorganisation of individual penguins is not represented. Figure \ref{fig:waters} shows results for various Péclet values and initial starting shapes of the huddle. In general, these show good agreement to the experiments of \cite{waters2012modeling}: the huddle propagates in the windward direction and evolves into a more streamlined shape over time. For large time ($t\gtrapprox t_{max}/2$), the huddle approaches an invariant egg-shape which continues to propagate downwind with some constant velocity. To test this observation, a root mean-squared error (RMSE) measuring the deviation between consecutive plotted huddle shapes is computed for the examples in Figure \ref{fig:waters} and found to approach a constant value with a relative error of $\mathcal{O}(10^{-4})$. This permits the reasonable conclusion that huddles do evolve into a steady shape and thus in all experiments presented, the maximum time $t_{max}$ is chosen such that the penguin huddle has reached a steady shape prior to or at time $t_{max}$.

\begin{figure}[h]%
\centering
\includegraphics[width=0.95\textwidth]{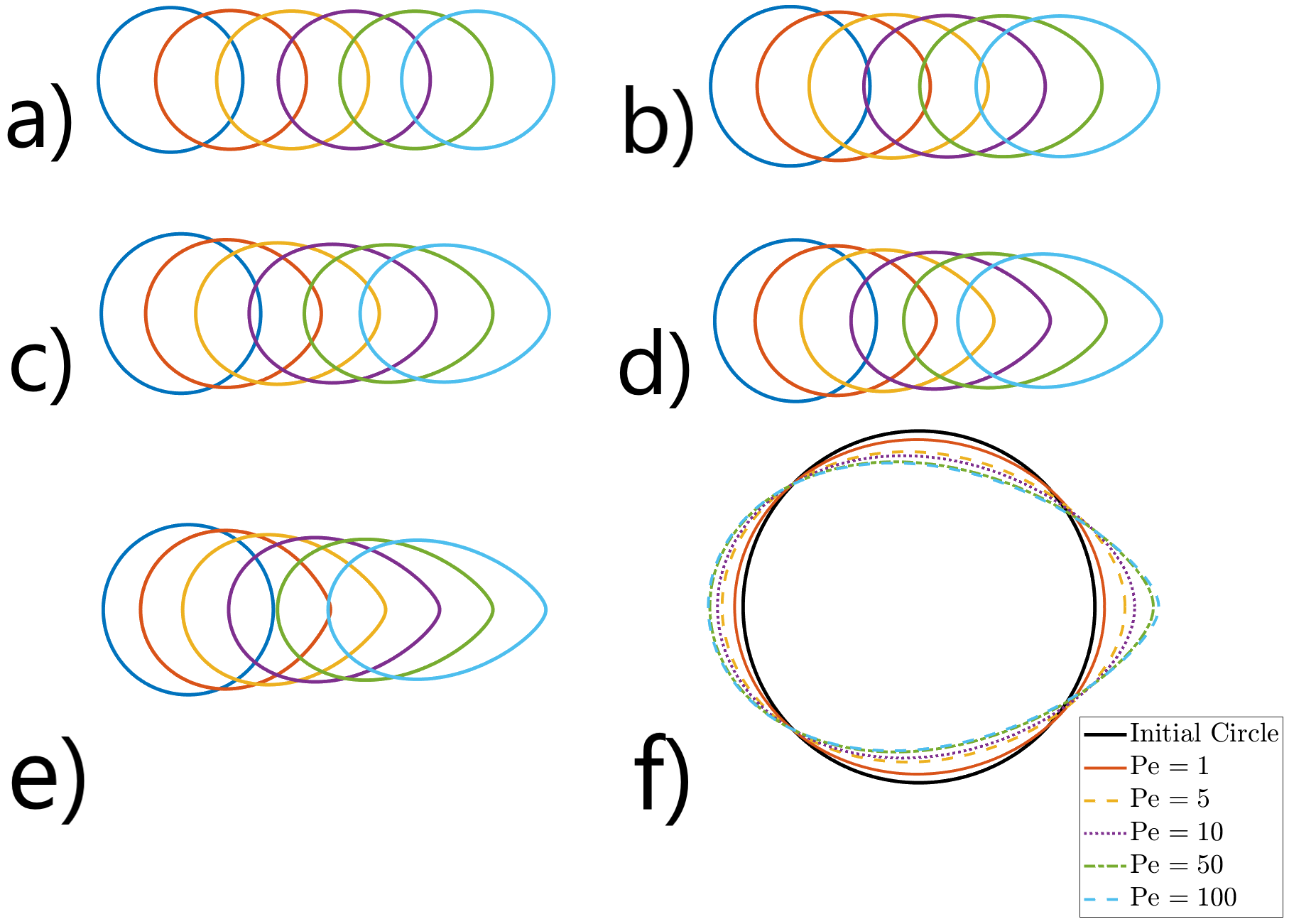}
\caption{The effect of different Péclet numbers on the evolution of the huddle is shown by plotting its shape at equal time intervals. The initial starting shape of the penguin huddle is a circle, with $\beta=0.1$ and a) Pe $=1$, $t_{max}=10$, b) Pe $=5$, $t_{max}=15$, c) Pe $=10$, $t_{max}=20$, d) Pe $=50$, $t_{max}=40$, e) Pe $=100$, $t_{max}=50$. f) Comparison of the steady shapes of (a)-(e), the initial (circular) huddle shape is also plotted for comparison.}\label{fig:diffPe}
\end{figure}

Now, consider the full penguin problem where exterior, interior ($\beta\neq0$) and area conserving terms are all included - recall from Sect. \ref{subsec:nondim} that $\beta=0.1$ is used. \textcolor{black}{Figures \ref{fig:diffPe}\textcolor{black}{a-e} present the evolution of the same initial starting shape, a circle, with Pe $=1$, $5$, $10$, $50$ and $100$. The final plot in each experiment is the steady shape and these are compared in Figure \ref{fig:diffPe}\textcolor{black}{f}, where the shapes have been re-centered so that they share the same centre of mass.} As the P\'eclet number is increased, the steady shape becomes less circular and more egg like.

\begin{figure}[h]%
\centering
\includegraphics[width=0.9\textwidth]{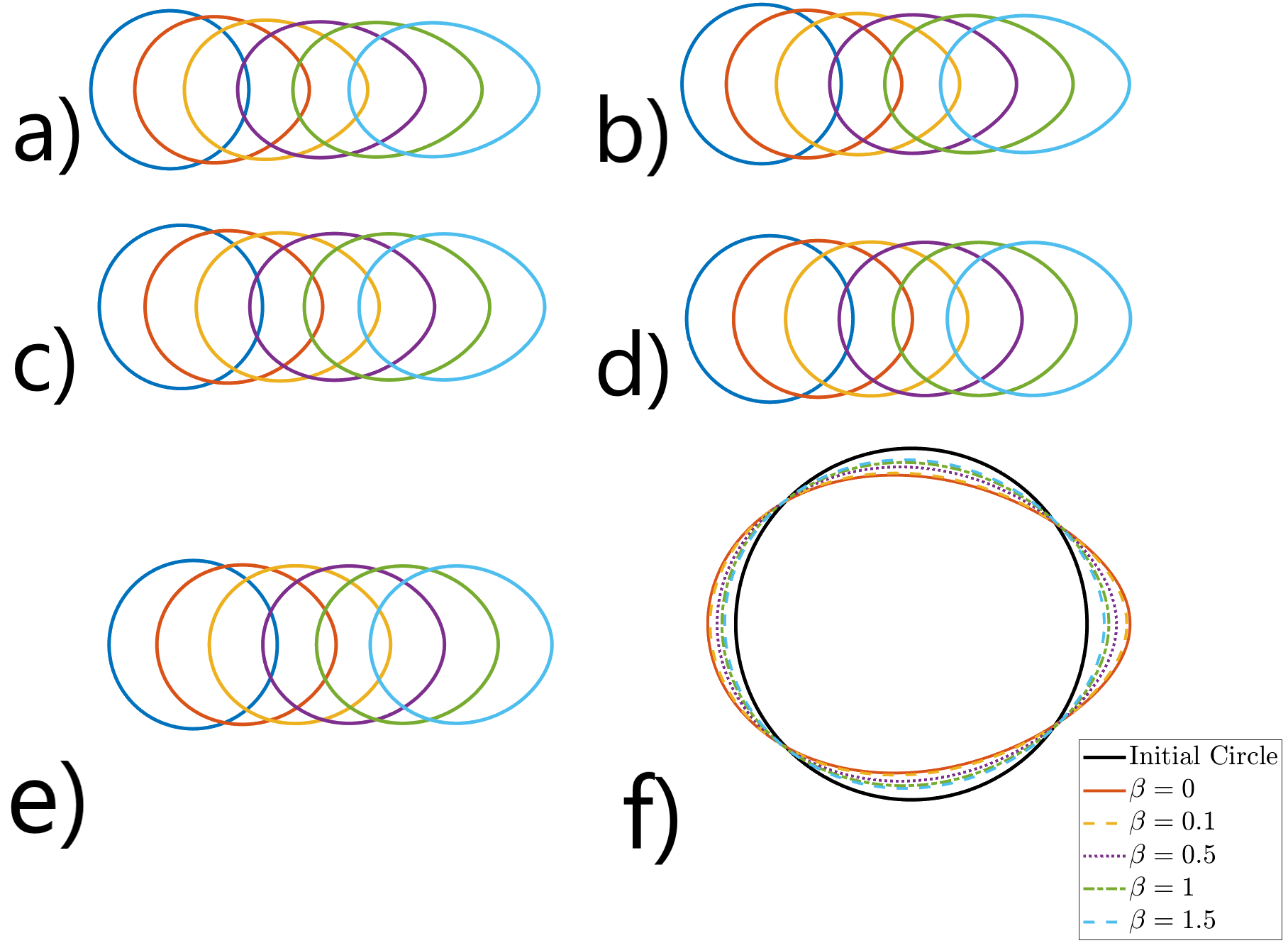}
\caption{Interior $\beta$ effect: initial starting shape of a circle with Pe $=10$, $t_{max}=20$ and a) $\beta=0$, b) $\beta=0.1$, c) $\beta=0.5$, d) $\beta=1$, e) $\beta=1.5$. f) Comparison of the steady shapes of (a)-(e), the initial (circular) huddle shape is also plotted for comparison.}\label{fig:diffbeta}
\end{figure}

\textcolor{black}{Similarly, the effect of changing the parameter $\beta$ can be explored. Figures \ref{fig:diffbeta}\textcolor{black}{a-e} shows the evolution of the circle with the values $\beta=0$, $0.1$, $0.5$, $1$ and $1.5$, respectively, where Pe $=10$ and $t_{max}=20$ are taken for all plots. 
While a low value of $\beta$ gives results that look similar to the exterior only penguin problem, a high value of $\beta$ causes the penguin huddle to retain a more circular shape. This is shown in Figure \ref{fig:diffbeta}\textcolor{black}{f}, where the steady shapes of Figures \ref{fig:diffbeta}\textcolor{black}{a-e} are compared. }

\begin{figure}[h]%
\centering
\includegraphics[width=0.97\textwidth]{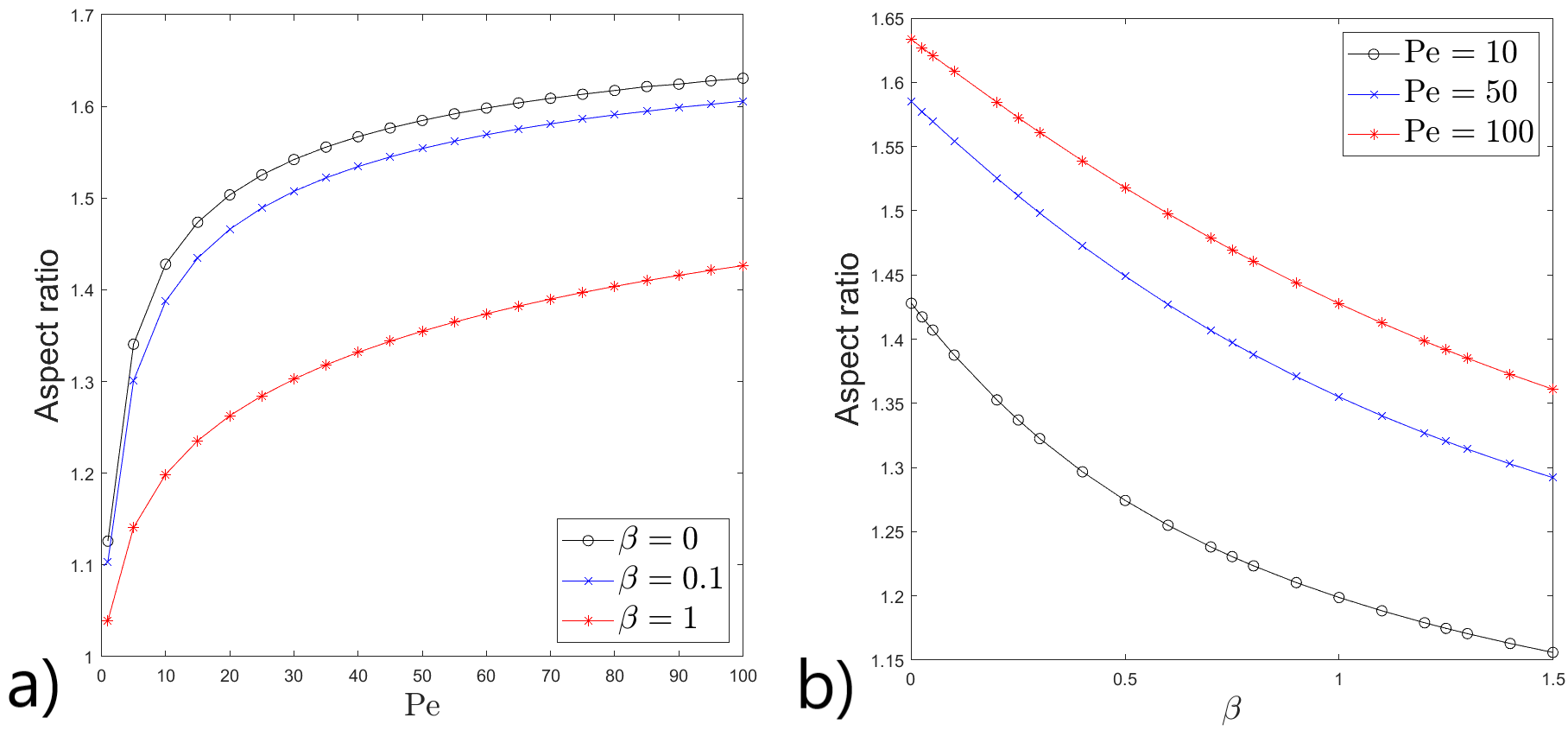}
\caption{The aspect ratio of the final steady shape of the penguin huddle as a function of a) P\'eclet number, for the three given values $\beta=0,0.1$ and $1$, and b) $\beta$, for the three given values Pe $=10,50$ and $100$. Each marker represents a numerical experiment.}\label{fig:arvspar}
\end{figure}

\textcolor{black}{The dependence of the final steady shape of the  huddle on the values of the P\'eclet number and $\beta$ is explored further. A given steady shape can be quantified by its aspect ratio (AR): the ratio between its major and minor axes with AR $=1$ being a circle and increasing ${\rm AR}>1$ becoming more egg-shaped. Figures \ref{fig:arvspar}\textcolor{black}{a,b} show the dependence of AR on Pe and $\beta$, respectively. Each marker on the diagram represents an experiment run: each curve contains 21 markers, all run to a maximum time of $t_{max}=80$, which is sufficiently long to reach the steady state huddle shape.} 

\textcolor{black}{The results from Figure \ref{fig:arvspar} agree with the conclusions drawn from Figures \ref{fig:diffPe} and \ref{fig:diffbeta}. First, as Pe increases,  AR  increases and thus a higher P\'eclet number results in a more egg-like steady shape, with a lower value of $\beta$ giving higher values of AR for all time. This is also consistent with the parameter study in \cite{waters2012modeling} - see their Figure 3 - which shows that the huddle thickness decreases with an increasing P\'eclet number. As Pe $\rightarrow 0$, AR$\rightarrow 1$, consistent with the limit of a circular steady shape when there is no wind, as expected. Second, as $\beta$ increases, AR decreases, where lower values of Pe result in lower values of the AR for all time. }

\begin{figure}[h]%
\centering
\includegraphics[width=0.92\textwidth]{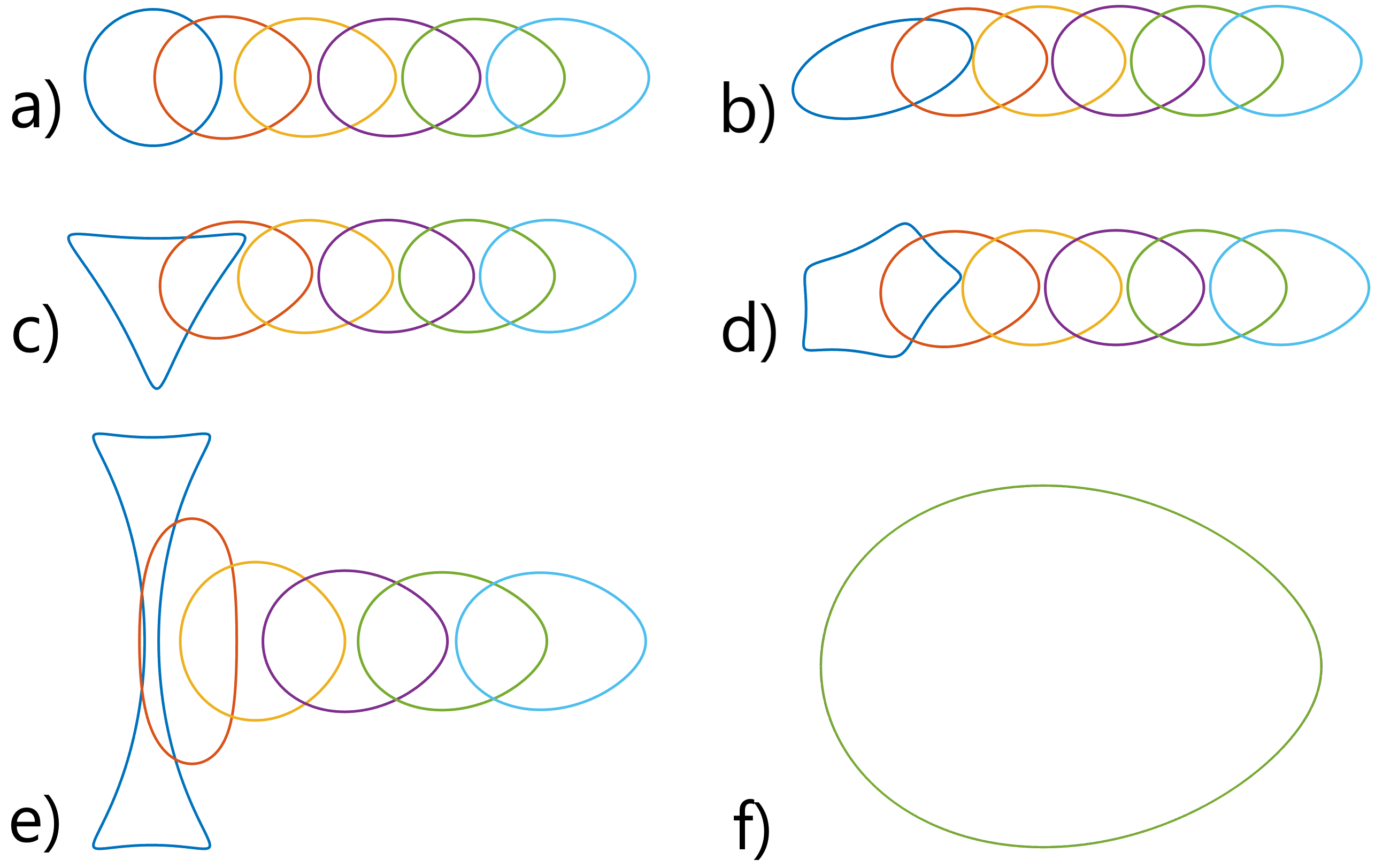}
\caption{Initial shape effect. Shapes are plotted at equal time intervals with Pe $=10$, $\beta=0.1$, $t_{max}=35$ and the initial starting shape of a) a circle, b) a slanted ellipse, c) a triangle, d) an irregular pentagon, e) an hourglass. f) Comparison of the steady shapes of (a)-(e): \textcolor{black}{all of the steady shapes are overlapped.}}\label{fig:diffshapes}
\end{figure}

Figure \ref{fig:diffshapes} shows how huddles of different starting shapes evolve under the same P\'eclet number $\text{Pe}=10$, \textcolor{black}{the same interior effect $\beta=0.1$ and up to the same maximum time $t_{max}=35$. Figures \ref{fig:diffshapes}\textcolor{black}{a-e} show the evolution of a circle, a slanted ellipse, a triangle, an irregular pentagon and an hourglass, respectively, and Figure \ref{fig:diffshapes}\textcolor{black}{f} compares the steady shapes of these five experiments.} The results show that, irrespective of the starting shape of the huddle \textcolor{black}{and even those starting with significant concave boundaries}, the same steady shape is reached. 

\begin{figure}[h]%
\centering
\includegraphics[width=0.9\textwidth]{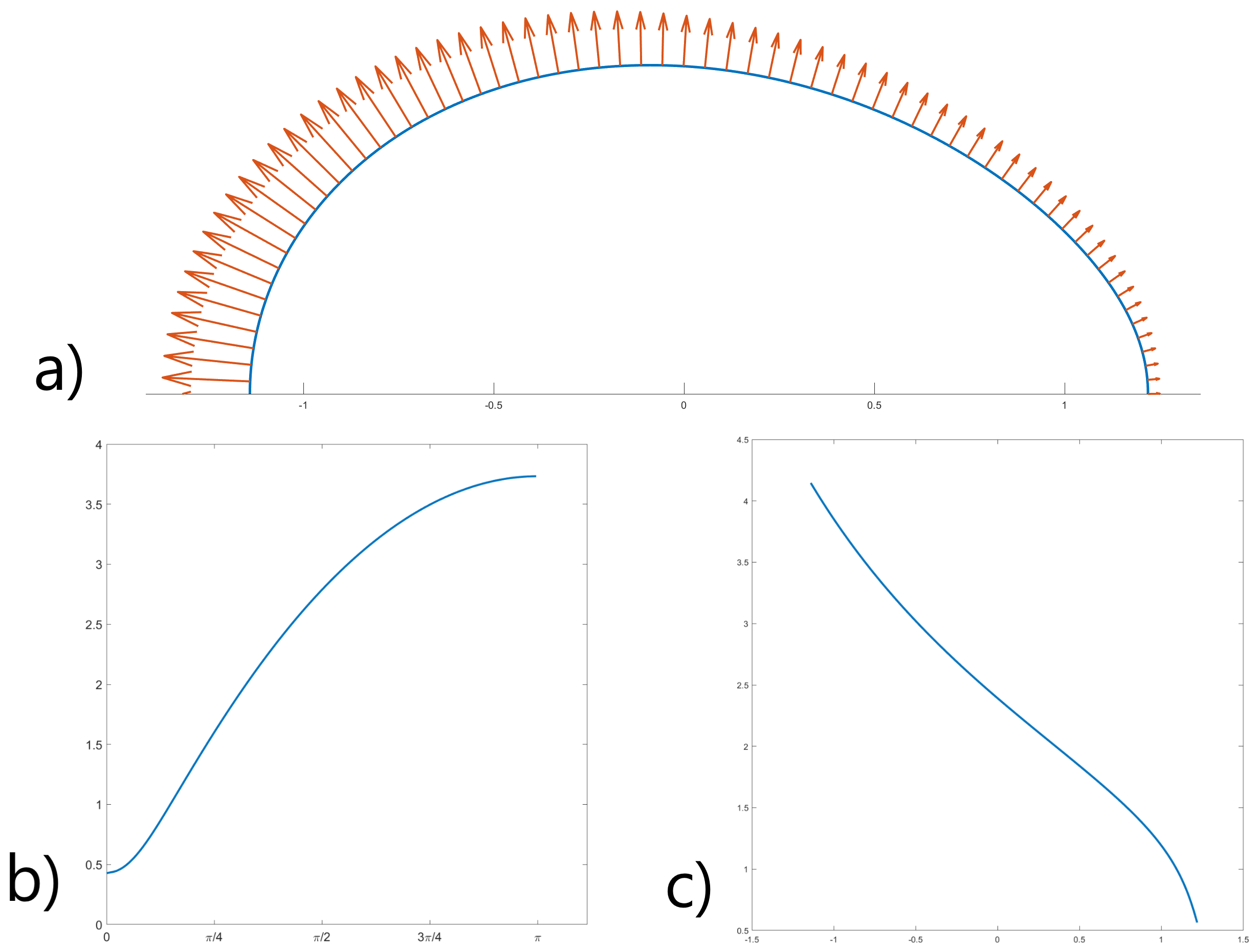}
\caption{a) Total heat flux across the boundary of the steady shape of Figure \ref{fig:diffshapes}\textcolor{black}{f}. The arrows point in the normal direction of the given boundary point and their magnitude is proportional to the local heat flux across the boundary. b) Canonical heat flux in the upper half $\zeta-$disk vs $\theta$. c) Physical heat flux in the upper half of the huddle boundary vs $x$.}\label{fig:heatflux}
\end{figure}

The along-boundary variation of the total heat flux across the huddle boundary of the steady shape of \textcolor{black}{Figure \ref{fig:diffshapes}\textcolor{black}{f}} is considered. By symmetry, only the upper half of the boundary need be considered; this is plotted in Figure \ref{fig:heatflux}\textcolor{black}{a}. The arrows point in the normal direction of the given point on the boundary and their lengths represent the total physical heat flux at that point. Recall from Sect. \ref{sec:num} that the total heat flux in the physical domain across the huddle boundary $\gamma$ is given by $\boldsymbol{\hat{n}\cdot\nabla}T_\Omega-\beta\boldsymbol{\hat{n}\cdot\nabla}T_R$ whereas the total heat flux in the canonical domain across the unit $\zeta-$disk is given by $\sigma(\theta)-\beta\omega(\theta)$. The canonical heat flux in the upper half $\zeta-$disk vs $\theta$ is plotted in Figure \ref{fig:heatflux}\textcolor{black}{b} and the physical heat flux vs $x$ is plotted in Figure \ref{fig:heatflux}\textcolor{black}{c}. There is a much larger heat flux on the windward side of the penguin huddle (the side of the penguin huddle directly facing the oncoming wind) compared to the leeward side. This is expected and is in agreement with the exterior temperature profile presented in \cite{waters2012modeling} - see their Figure 1 - which shows that there is a steeper temperature gradient, and hence a higher total heat flux, on the windward side.

\textcolor{black}{It seems likely that  there is randomness associated with penguin movements owing to, for example, inhomogeneities in the thermal properties of penguins on the boundary, or fluctuating wind in the exterior.  Waters {\it et al.} \cite{waters2012modeling} incorporate time-varying perturbations by randomly varying the heat loss of boundary penguins at each time-step. Owing to the deterministic ODE solvers used in the numerical model here, such time-dependent random perturbations are not tackled in this work. However, as a step towards considering random effects, a steady random perturbation to the exterior heat flux from the penguin huddle to the wind is considered using a modified version of \eqref{Meq:PGvn}}

\begin{equation}\label{Meq:stochPG}
    \text{Pe Re}[f_t\overline{\zeta f_\zeta}] = (1+\epsilon\mu)\sigma(\theta)-\beta\omega(\theta)+C(t)\lvert f_\zeta\vert,
\end{equation}
\textcolor{black}{where $0<\epsilon\ll1$ is some constant and, when the equation is discretised in $\theta$, $\mu(\theta_j)=\mu_j$, $j=1,\cdots,2N+3$, are random numbers selected from a uniform distribution between $-1$ and $1$.}

\begin{figure}[h]%
\centering
\includegraphics[width=0.75\textwidth]{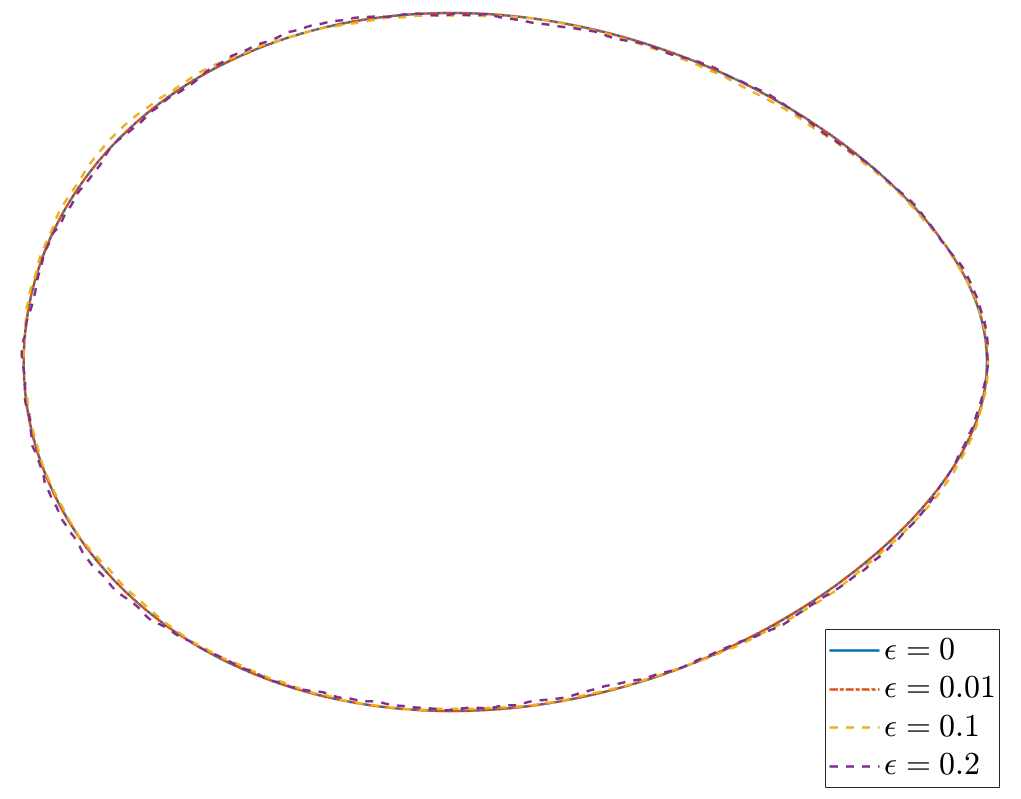}
\caption{Comparison of the steady shapes of the propagation of a circular penguin huddle with Pe $=10$, $\beta=0.1$, $t_{max}=20$ and different amplitudes $\epsilon$ of the random function $\mu(\theta)$: $\epsilon=0$ (no randomness), $\epsilon=0.01$, $\epsilon=0.1$ and $\epsilon=0.2$.}\label{fig:random}
\end{figure}

\textcolor{black}{Figure \ref{fig:random} compares the steady shapes of four experiments, each simulating the propagation of an initially circular penguin huddle with Pe $=10$, $\beta=0.1$, $t_{max}=20$ and different amplitudes $\epsilon$ of the random function $\mu(\theta)$: $\epsilon=0$ (no randomness), $\epsilon=0.01$, $\epsilon=0.1$ and $\epsilon=0.2$. The $\epsilon=0.01$ experiment shows no discernible difference between the $\epsilon=0$ experiment, and while there are some slight variation between the $\epsilon=0.1,0.2$ and $\epsilon=0$ experiments, these are very minor. All experiments result in the same general egg-shaped steady shape which, along with those of Figure \ref{fig:diffshapes}, demonstrate the persistence of the steady egg-shaped huddles.}

\section{Discussion}\label{sec:discuss}

In this continuum model, the evolution of the penguin huddle boundary $\gamma(t)$, is affected by three factors: exterior cooling by the ambient wind, interior reorganisation of the penguins driven by their self-generation of heat, and area conservation of the huddle. The resulting two-dimensional model \eqref{Meq:vn}-\eqref{Meq:C} can be rewritten as a PG-type equation \eqref{Meq:PGvn} in terms of the conformal map $z=f(\zeta,t)$ between  the canonical ($\zeta$) and physical ($z$) planes. The map is approximated as a truncated Laurent series \eqref{eq:cmap}, whose unknown time-varying coefficients are found by numerically solving a system of coupled ODEs. Results from Sect. \ref{sec:results} show that the huddle shape eventually reaches a steady shape. As the wind strength increases, the huddle becomes increasingly egg-shaped, minimising the portion of the huddle boundary directly facing the oncoming wind. For a stronger interior heating effect, the huddle approaches a more circle-like steady shape, to optimise interior heat regulation. For given wind and interior effects, any initial huddle shape eventually approaches the same  steady shape. There is also a greater total heat flux on the windward side of the huddle than on the leeward side.

While dimensionless quantities have been used throughout the paper,  it is useful to estimate the  dimensional time scale associated with the huddle evolution. \textcolor{black}{As noted in Sect. \ref{subsec:nondim}, the appropriate dimensional time scale is $t_r=\lambda\text{Pe}(L/U)t^\ast$ where $t^\ast$ is dimensionless time, and the choice $\lambda=1$ has been made.} A large huddle of, for example, 6000 penguins \cite{le1977emperor}, is assumed and taking the radius of each penguin to be $r\approx0.1$m \cite{williams2015hidden} gives $L\approx10$m. Now assume a sufficiently strong wind of $U\approx20$m$s^{-1}$ for Pe $=100$. Considering Figure \ref{fig:diffPe}\textcolor{black}{c} where $t^\ast=t_{max}=50$, the total real time for this experiment is $t_r=2500$ seconds  or $\approx 42$ minutes. Thus it is concluded that the penguin huddle motion described in this work occurs over roughly a one hour period, in agreement with previous modelling and field observations \cite{gilbert2006huddling,waters2012modeling,kirkwood1999occurrence}, \textcolor{black}{and vindicating the choice $\lambda=1$.}

In this paper, a continuum model has been used to find the shape and propagation of the huddle boundary at the expense of tracking individual penguin movements. While there are pre-existing discrete models for penguin huddling (and other animal grouping dynamics) \cite{waters2012modeling,gerum2013origin,gerum2018structural,bernardi2020agent}, treating a penguin huddle as a continuum appears to be new. \textcolor{black}{One of the advantages of the continuum model developed here is its computational efficiency, with a typical runtime of around 5-10 minutes on a standard laptop for a huddle to reach a steady shape. This can be sped up further, to a matter of seconds, by reducing the Laurent series truncation $N$, though care should be taken to ensure that the accuracy of the results is not compromised.}

When considering the ``exterior only" ($\beta=0$) penguin problem, our results matched well with those from \cite{waters2012modeling}, who did not include the interior effect of penguin reorganisation in their model. \textcolor{black}{As discussed in Sect. \ref{subsec:nondim}, fully turbulent wind flows have $\beta\approx0$ and hence are well modelled by considering exterior effects alone. Realistic choices for Antarctic conditions have  $\beta\approx 0.1$, in which case Figure \ref{fig:diffbeta}\textcolor{black}{f} demonstrates that the interior heating of the colony only provides a weak effect in determining the steady shape of the huddle. Larger values of $\beta$ resulting in more circular huddle shapes have been included here for completeness. Further, the results in this work match well with real world observations: penguin huddles are seen to march downwind over time and observed huddle shapes resemble the egg-like steady shapes found here - see \cite{le1977emperor,richter2018phase,gerum2013origin,mina2018penguin,ancel2015new}.}

Extensions to the penguin problem are of interest. A time-varying wind could be considered, where the variations are on the order of an hour; when the wind direction changes, the huddle will seek to propagate back to the optimal (egg) steady shape in the new windward direction. \textcolor{black}{As in \cite{waters2012modeling} and noted in Sect. \ref{sec:results}, stochastic terms could also be incorporated into the model. This includes random variations in the direction and magnitude of the wind and stochasticity in the interior and exterior heat fluxes. The numerical method would need to be adapted to handle these stochastic terms.}

The evolution of any arbitrary starting huddle shape could also be simulated. In this work, the initial huddle boundary was taken such that its conformal map was readily represented in  the form of a truncated Laurent series \eqref{eq:cmap}. For more arbitrary starting shapes, the huddle boundary can be discretised as a polygon of $m$ vertices, then its conformal map to the $\zeta-$plane found numerically at each time step as the penguin huddle evolves. Pre-exisiting numerical schemes for finding the conformal map between a polygon and the unit $\zeta-$disk include the Schwarz-Christoffel toolbox \cite{driscoll1996algorithm} and a variation of the AAA-LS algorithm \cite{costa2021aaa} from Sect. \ref{sec:num}. \textcolor{black}{It would also be possible to consider multi-connected huddles such as those with penguin-free holes or the interaction, and possible merger, of initially disjoint huddles. While this would require a significant modification of the numerical method used here, it is worth remarking that  AAA-LS type methods are able to handle such geometries \cite{costa2021aaa,costa2020solving,trefethen2020numerical}}.

Finally, consider the boundary condition \eqref{Meq:vn}, and its associated PG equation \eqref{Meq:PGvn}. The right-hand side is constructed by exterior ($\boldsymbol{\hat{n}\cdot\nabla}T_\Omega$), interior ($\beta\boldsymbol{\hat{n}\cdot\nabla}T_R$) and area conservation (C(t)) effects. By inclusion and exclusion of these three terms, \eqref{Meq:vn} encapsulates six possible free boundary problems:
\begin{enumerate}
\item The dissolution problem: exterior effect only. A permeable object is placed in a uniform stream of some dissolving agent - see \cite{dutka2020time,ladd2020}.
\item The Poisson growth problem: interior effect only. Applicable to squeeze flow and the evaporation of thin liquid films - see \cite{agam2009viscous,crowdy2001squeeze,mcdonald2015poisson}.
\item The two-phase melting/freezing problem: exterior and interior effects. Porous media flow about freeze pipes where interior and exterior temperature gradients are in effect - see \cite{goldstein1978effect,rycroft2016asymmetric}.
\item The ``exterior only" penguin problem: exterior and area conservation effects. Huddle problems where interior reorganisation is not represented, such as if individuals in the centre are unable to move or there exists a ``pecking order" in the population - see \cite{waters2012modeling}.
\item The ``bat huddle" problem: interior and area conservation effects. Huddle problems with no wind effect, for example bats huddling in caves sheltered from exterior winds. \textcolor{black}{To the authors' knowledge}, there is no mathematical work on this problem, \textcolor{black}{but for biological context see} \cite{herreid1963temperature,ryan2019changes}.
\item The ``full" penguin problem: exterior, interior and area conservation effects. This is the full huddle continuum problem covered in this work.
\end{enumerate}
Therefore, the numerical method developed in this work can be used to model and simulate additional free boundary problems, such as those from the fields of fluid mechanics and mathematical biology.

\section*{Acknowledgments}

The authors thank Carole Hall, Stony Brook University, for their inspiring talk on Ad\'elie penguins and the useful discussion thereafter. \textcolor{black}{The authors also thank the anonymous reviewers for their useful comments and suggestions.}

\section*{Statements and Declarations}

\begin{itemize}
\item \textbf{Funding:} Samuel J. Harris was supported by a UK Engineering and Physical Sciences Research Council PhD
studentship, grant numbers EP/N509577/1 and EP/T517793/1.
\item \textbf{Data availability:} All data generated or analysed during this study are included in this published article and its supplementary information files.
\item \textbf{Code availability:} All relevant code files are publicly available at the following GitHub link: https://github.com/Sam-J-Harris/Penguin-huddling-a-continuum-model.git.
\item \textbf{Competing interests:} Not applicable.
\end{itemize}

\bibliographystyle{unsrt}  
\bibliography{AB-bibliography}

\end{document}